# INDUSTRIAL APPLICATION OF NUMERICAL MODELS FOR ALUMINIUM EXTRUSION

Juan M. Torres Zanardi†, Ana Scarabino†, Federico Bacchi† and Luciano Principi‡

*Computational Fluid Dynamics Group - GFC, Universidad Nacional de La Plata (UNLP), 116 Street between 47 & 48, La Plata, Argentina, gfc@ing.unlp.edu.ar, www.gfc.ing.unlp.edu.ar*
*‡MADEXA S.A., 182 No. 575, Lisandro Olmos, Argentina, info@madexa.com.ar, www.madexa.com.ar*

ABSTRACT: This study presents the numerical models used for the simulation of the large viscoplastic deformations that aluminium undergoes during the extrusion process in order to obtain industrial profiles. This study also gives examples of results obtained by the Computational Fluid Dynamics Group of the UNLP Faculty of Engineering, in collaboration with the company Madexa S.A., dedicated to the manufacturing of dies for this type of processes. The equations that model the process, the difficulties associated with its numerical resolution and the advantages that the simulation work represent for the company are also presented in this study.

## 1. INTRODUCTION

Extrusion is a plastic deformation process in which a metal block (billet) is forced by compression to flow through an opening of a smaller cross-section. This produces one or more profiles with the shape determined by the output cross-section. This process can be done cold (for example, for wires) or, more frequently, hot, when the billet is preheated before being introduced into the die. The designs of the dies should ensure that the material flows through the exit section with as uniform a velocity as possible, in order to avoid curved or warped profiles on leaving the die. At the same time, the die must be of a high structural stiffness so as not to greatly distort nor break because of the high stresses exerted during the process [1].

The ratio between the aluminium billet cross sections and the extrusion profiles can be as large as 200 to 1 or even larger. The large plastic deformations undergone by the material during the extrusion create significant local differences in shear, strain rate and heat generation. These differences make the objective of designing a die that produces a uniform material output velocity a considerable challenge, especially in view of the myriad of profile cross-sections required by the industry: open, tubular, mixed, with different thicknesses, etc. Figure 1 presents a schematic of an extrusion system and some examples of profiles produced by this process. The die design that achieves our objectives is heavily based on experience and is frequently achieved through a process of trial-and-error: if generated profiles from a first die have no satisfactory results, the corrections to be applied to it are determined, or, failing that, to a second die, and so on. This process entails production delays and economic losses for both the die manufacturer and the profile manufacturer.

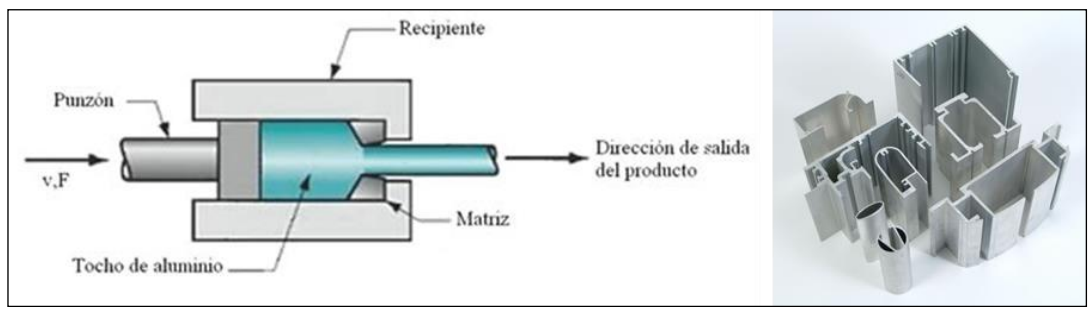

Figure 1: Left, extrusion die schematic (taken from Galmak Alüminyum.) Right, different types of profiles (GSH Industries, Inc.)

In this context, the possibility to perform a numerical prediction of the profile formation, detecting possible problems before beginning the die construction, grants the opportunity to reduce the redesign time and to save the material, man-hours and machine-hours spent to modify the existing die or rather, the design and manufacturing of a new die to correct necessary details to obtain a homogeneous extrusion. Typically, this redesign process, arising from the detection of the need for changes in a die, requires at least 3 to 5 engineering and production man-hours plus the use of materials and machinery for the die manufacturer. Additionally, the production start for the profile manufacturer requires 48-72 hours of delay resulting in economic losses for both.

At the same time, applying the Finite Element method (FEM) to the extrusion processes makes it possible to identify the influence of different parameters (such as billet initial temperature or ram speed) in variables such as the loads the die is subjected to, material temperatures, stress distribution and material strain rates, etc. This method proves to a be a valuable research tool, as it has been demonstrated in numerous publications [2,3].

The aim of this work is to present examples of the improvement of extrusion quality obtained through the application of numerical simulation.

## 2. METHODOLOGY

For the analysis of stress, deformation and temperature inside the material during extrusion, several models are available. A standard model widely used and rated is to consider the material behaviour during hot extrusion similarly to that of a non-Newtonian fluid in which the apparent viscosity is a function of the strain rate and temperature [3]. This model is applied in software packages specifically intended to simulate extrusion processes (both for metals and plastics and other materials) in different commercial programs, such as Ansys® Polyflow, Altair® Inspire Extrude or QForm®, and it is the one used in the results of this work.

The equations to be solved by the software are those of conservation of mass, momentum for a creeping flow (Stokes) and energy, taking into account their coupling through the heat generation by deformation work and viscosity change associated to temperature and strain rate changes.

$$\nabla \cdot \mathbf{V} = 0$$
$$\nabla \cdot \left( -P\mathbf{I} + \boldsymbol{\sigma}(\dot{\gamma}, T) \right) = 0$$
$$\rho C_p(T) \frac{\partial T}{\partial t} = -\rho (\mathbf{V} \cdot \nabla) T + \nabla \cdot (k \nabla T) + \boldsymbol{\sigma}(\dot{\gamma}, T) : \mathbf{D}$$

In these equations, the $\mathbf{D}$, $P$, $\mathbf{I}$, $\boldsymbol{\sigma}$, $\mathbf{V}$, $T$, $\dot{\gamma}$, $\rho$ and $C_p$ quantities stand for, respectively, the strain rate tensor, pressure, identity tensor, stress deviator tensor, flow velocity vector, temperature, deformation rate, density and specific heat at constant pressure. The stresses are function of the temperature and magnitude of $\dot{\gamma}$, defined as:

$$\dot{\gamma} = \left( \frac{2}{3} D_{ij} D_{ij} \right)^{1/2}, \quad \text{with} \quad \mathbf{D} = \frac{1}{2} \left( \nabla \mathbf{V} + \nabla \mathbf{V}^T \right)$$

The relation between stresses, strain rate and temperature is obtained by fitting experimental data equations. A commonly used model is the "Inverse Hyperbolic Sine" model, the parameters of which are defined in the reference [4].

$$\bar{\sigma} = \frac{1}{\alpha} \sinh^{-1} \left[ \left( \frac{1}{A} \dot{\gamma} e^{\left( \frac{Q}{RT} \right)} \right)^{\frac{1}{n}} \right]$$

Figure 2 shows the non-linearity relation between stress, strain rate and temperature for an AA6063 aluminium, which responds adequately to this model.

Even if, due to the high viscosities, the convective acceleration terms are neglected in the momentum equation (Stokes flow), the set of equations to be solved is highly non-linear and requires iterative formulations that demand considerable computational time.

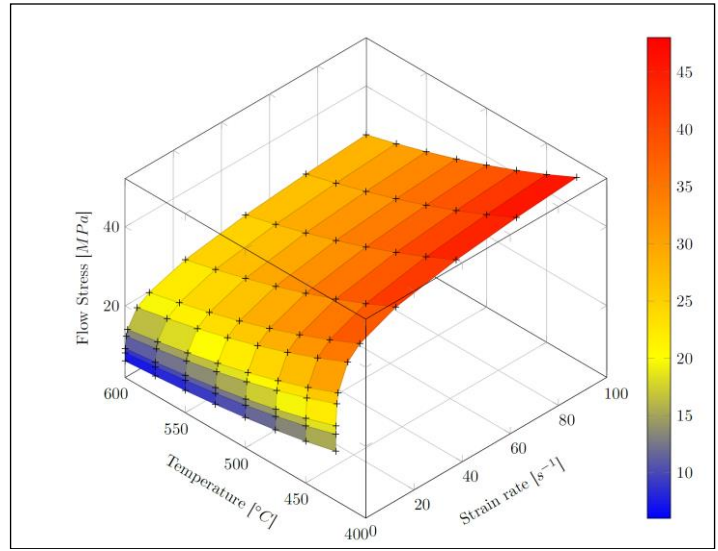

Figure 2: Relation between stress, strain rate and temperature for AA6063 aluminium

The methodology of work between the different actors consists of the following steps:
- Based on the client's order (aluminium profiles manufacturing company), the die manufacturer generates the design and CAD file of the die for the desired profile. The client establishes parameters such as the aluminium billet and container diameter, the container pre-heating temperature, the press capacity that pushes the ram, the ram speed, etc.
- The company provides the CAD file. Since this CAD is prepared for production and not for simulation, the geometry goes through a debugging process to remove details that may cause numerical errors or convergence problems.
- A hybrid mesh is generated with tetrahedral elements, except at the profile exit, where prisms aligned with the extrusion direction are used.
- The solver is set up establishing the material constitutive relations.
- The mass, momentum and energy conservation equations are solved with an Arbitrary-Lagrangean-Eulerian (ALE) approach [4].
- The results are post-processed, the main quality control element of the die being the uniformity of the material velocity through the output section and the absence or minimisation of lateral velocity components that produce a curvature of the profiles. This curvature causes the impossibility for profiles to fit in the blueprint geometry dimensions provided by the client, within prescribed tolerances as per the IRAM-699 standard. Also, in this phase, stresses on the die can be verified to not cause excessive deformations or risk of breakage, the force to be applied to the billet can be verified to not exceed the press capacity available, alongside additional results of potential interest, such as stress distribution, temperature, velocity field, etc.

A mesh suitable for extrusion simulation requires a number of elements in the order of 2-3 million, depending on the complexity of the die and profiles to be produced. Convergence for normalized residuals of 1e-4 tends to be achieved in about 30 non-linear iterations, leading to a total computation time of about 4 hours per million elements (i.e., typically 8-12 hours) on a Workstation with 2 Intel Xeon E5-2640 v3 CPUs, each with 8 physical cores, and 64 GB of RAM.

## 3. EXAMPLE OF DIE OPTIMIZATION

This goal is illustrated by the results obtained with the commercial Altair® Inspire Extrude software for a die designed to produce six identical profiles in each working cycle. In this case, production requirements dictated that all six profiles had to have the same horizontal orientation, so their radial distribution could not be periodic or symmetrical.

The results obtained with the first design were not satisfactory, since major speed differences between the different profiles could be seen, as well as locally in each exit. Based on these results, the die was completely redesigned, and a much more homogeneous output was achieved, as seen in figure 4b. Exit speed deviations were reduced from -72 % - +92% (with lateral components producing section warping and axis curvature in some of the profiles) to -3.7% - +6.6%, results within the range considered as very acceptable.

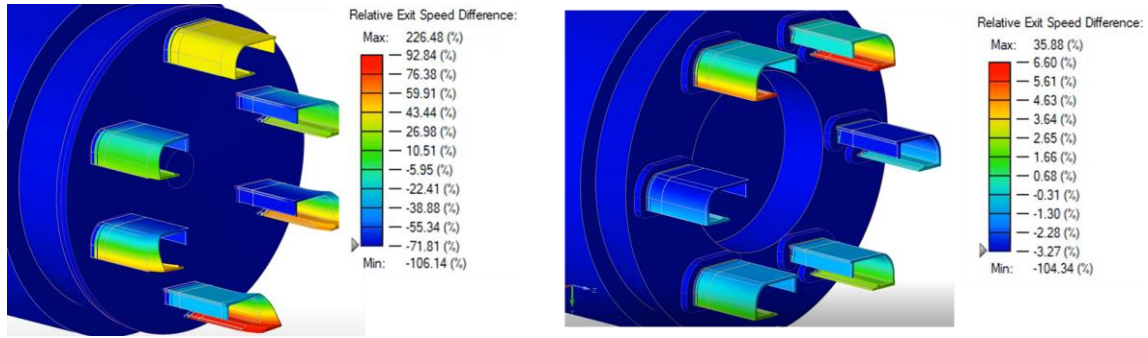

Figure 4: Die extrusion results for multiple profiles. Left: Original design. Right: Modified design based on the obtained results from the original design.

## 4. CONCLUSIONS

The numerical simulation of the aluminium profiles extrusion process has been described, as well as the mathematical complexities of modelling the metal subjected to large viscoplastic deformations. In a calculation example it has been shown how, based on numerical results, it was possible to redesign a die, which significantly improved the quality of profiles produced with it. Using numerical tools to evaluate and redesign dies has proved to be profitable for local SMEs, as it reduces costs and time associated with construction, delivery, testing and recovery of a die that requires redesigning. Additionally, the differential in the quality of its products obtained thanks to the simulation results in greater customer satisfaction and contributes to an innovative business image.

## ACKNOWLEDGEMENTS

This work is part of a cooperation between the company Madexa S.A. and the Computational Fluid Dynamics Group of the Faculty of Engineering of the Universidad Nacional de La Plata. We would like to thank Diego Principi, president of Madexa S.A., for his initiative to work in collaboration with us to explore the possibilities of Computational Fluid Mechanics for the simulation of aluminium profile extrusion, and for all the information provided on the technical and commercial aspects of this process. We are also thankful to Gabriel Pereira for his valuable help in the writing of this work.